\documentclass[aps,prl,preprint]{revtex4}
\usepackage{graphicx}
\usepackage{amsmath}
\usepackage{amssymb}
\usepackage{ulem}

\begin{document}

\baselineskip=1.5cm

\title{Inhomogeneous vortex-state-driven enhancement of superconductivity in nanoengineered ferromagnet-superconductor heterostructures.}

\author{R. K. Rakshit$^1$}
\author{R. C. Budhani$^{1}$}\email{rcb@iitk.ac.in}
\author{T. Bhuvana$^2$}
\author{V. N. Kulkarni$^3$}
\author{G. U. Kulkarni$^2$}

\affiliation{1. Condensed Matter - Low Dimensional Systems
Laboratory, Department of Physics, Indian Institute of Technology
Kanpur, Kanpur - 208016, India}

\affiliation{2. Jawaharlal Nehru Centre for Advanced Scientific
Research, Jakkur P.O., Bangalore 560 064, India}

\affiliation{3. Central Nuclear Laboratory, Department of Physics,
Indian Institute of Technology Kanpur, Kanpur - 208016, India}

\begin{abstract}

Thin film heterostructures provide a powerful means to study the
antagonism between superconductivity (SC) and ferromagnetism (FM).
One interesting issue in FM-SC hybrids which defies the notion of
antagonistic orders is the observation of magnetic field induced
superconductivity (FIS). Here we show that in systems where the FM
domains/islands produce spatial inhomogeneities of the SC order
parameter, the FIS can derive significant contribution from
different mobilities of the magnetic flux identified by two
distinct critical states in the inhomogeneous superconductor. Our
experiments on nanoengineered bilayers of ferromagnetic CoPt and
superconducting NbN where CoPt/NbN islands are separated by a
granular NbN, lend support to this alternative explanation of FIS
in certain class of FM-SC hybrids.
\end{abstract}

\maketitle
%
%
%


The pinning of magnetic flux lines in a superconducting (SC) film
by ferromagnetic (FM) dots and multidomain magnetic films has been
a topic of considerable interest
\cite{Lyuksyutov,Lange1,Igor,Buzdin,Gillijns,Pissas,Morgan1,Bekaert}.
An important consequence of the inhomogeneous magnetic field
produced by such structures is the formation of a local field-free
state when the FM-SC hybrid is placed in an external magnetic
field. This effect has been variously called as field-enhanced
superconductivity \cite{Lyuksyutov,Lange1}, magnetization
controlled superconductivity \cite{Igor} and domain wall
superconductivity \cite{Buzdin,Gillijns}.

Earlier studies of FM-SC hybrids have been carried out mostly on
films of elemental superconductor such as Nb and Pb deposited on
in-plane as well as high coercivity perpendicular anisotropy media
\cite{Lyuksyutov,Lange1,Igor,Buzdin,Gillijns,Pissas,Morgan1,Bekaert}.
Although striking field-induced changes in the critical
temperature (T$_c$) of the superconductor have been seen, the
field-free regions produced by the dipolar field of FM dots are
not homogeneous in these systems. Also, it is not clear whether
superconductivity actually survives in areas above (or below) the
dots because of the small upper critical field (H$_{c2}$) of the
elemental superconductors and strong magnetization of the dots.
The global superconducting order parameter in these FM-SC hybrids
is clearly inhomogeneous irrespective of the orientation of
magnetization in the dots. It is therefore, important to address
the possible role of granularity in causing the FIS in such
systems. The relevance of this issue becomes apparent from the
richness of the field-temperature (H-T) phase diagram of
well-known granular systems such as composite films
\cite{Vinokur}, high T$_c$ ceramics \cite{Tinkham} and
Josephson-junction arrays \cite{Martinoli}.


In this report we address field-induced superconductivity in
nanoengineered grids of CoPt/NbN where the CoPt-free regions have
been made granular, with a reduced T$_c$. The advantage of using
NbN is in its large H$_{c2}$ ($>$ 250 kOe) which prohibits a
significant perturbation of superconductivity by the CoPt. The
resistance (R) of the nanoengineered structures has been measured
with current density J $\geq$ J$_c$ as a function of external
field $\vec{H}$ applied in two different orientations; (i)
($\vec{H}$ $\perp$ $\vec{J}$) $\perp$ $\hat{n}$, and (ii),
($\vec{H}$ $\parallel$ $\vec{J}$) $\perp$ $\hat{n}$, where
$\hat{n}$ is a unit vector normal to the plane of the film. The R
vs. H loops are hysteretic with their reverse branch showing entry
into the superconducting state at fields much higher than the
field at which R appears in the forward branch. This FIS state is
shown to be a consequence of the granularity of magnetic and
superconducting order parameters in the hybrid.


Thin film bilayers of CoPt and NbN each 50 nm thick, were
deposited on (001) MgO at 600 $^0$C using pulsed laser ablation of
CoPt and Nb targets respectively. Further details of the
deposition of NbN and CoPt films are given elsewhere
\cite{Rakshit,Senapati,Budhani}. The bilayer film was first
patterned with standard lithography and then Ar$^+$ ion milled
into a 100 $\mu$m$\times$500 $\mu$m bridge as shown in Fig. 1. The
top CoPt layer was then structured in simple square patterns using
a Ga$^{3+}$ focused ion beam (FIB) milling facility. Fig. 1(b)
shows a typical scanning electron micrograph of the patterned
bilayer. The size of each CoPt element in the pattern is
0.5$\times$0.5 $\mu$m$^2$ with 50 nm spacing in between. The
active area of the sample on which transport measurements were
carried out contains 10$\times$210 square elements.


The magnetic domain structure of the CoPt squares was probed with
Magnetic Force Microscopy (MFM) and Atomic Force Microscopy (AFM)
performed simultaneously on a thermally demagnetized sample in the
tapping mode at various lift heights. A commercial magnetometer
was used for measurement of magnetization. The critical current
density (J$_c$), resistance R (T) and magnetoresistance R (H) were
measured between 5 and 20 K.

Since the central result of this paper is based on inhomogeneities
of the SC and FM order parameters, we first establish these
inhomogeneities in the patterned sample. Fig. 2 presents R (T) of
the patterned as well as unpatterned bilayer. While the latter
film shows a sharp transition ($\Delta$T $\approx$ 0.5 K) with
T$_c$ onset at $\approx$ 15.3 K, the R (T) of the patterned film
exhibits a two step structure - where the first drop in resistance
occurs at $\approx$ 15.1 K in which R decreases by $\approx$ 60
$\%$ of its normal state value. The second transition starts at
$\approx$ 13 K, culminating in R = 0 at $\approx$ 11.8 K. The
change from single-step to two-step transition on nanostructuring
is presumably a result of Ga$^{3+}$ ion damage. This feature,
however, provides the unique advantage of addressing FIS in a
granular medium. Inset of Fig. 2 compares the J$_c$ (T) of the
nanostructured and a plane bilayer film. The J$_c$ of the
unpatterned film quickly reaches the limit of our measurement
which is set by the width of the bridge and maximum current used
($\approx$ 100 mA). However, a pronounced reduction in the J$_c$
is seen on nanopatterning. Since the capacity of such structures
to carry dissipation-less current is determined by their low T$_c$
links, a suppressed growth of J$_c$ on lowering the temperature
indicates that the channels of NbN connecting CoPt/NbN islands are
granular.


Fig. 2 also shows magnetization (M) of an unpatterned bilayer at 5
and 20 K as a function of magnetic field applied along the surface
of the film. The square hysteresis loop of low coercivity
($\approx$ 250 Oe) at 20 K seen here suggests a soft ferromagnetic
state in CoPt with in-plane $\vec{M}$, a characteristic feature of
the disordered fcc phase of CoPt \cite{Budhani}. The M-H curve
traced at 5 K is dominated by the diamagnetic response of the NbN
layer. While it was not possible to measure the magnetic state of
the patterned film with SQUID magnetometry, we have used room
temperature MFM to visualize domain structure of the CoPt islands.
Fig. 1(c) and (d) show the AFM and MFM micrographs respectively,
taken from the same spot on the film \cite{Text}. The bright and
dark features in MFM represent domains of reversed $\vec{M}$.
Since each patterned CoPt island has two such domains, it is clear
that they are in a thermally demagnetized state.


In Fig. 3, we show a set of R (H) data for the nanostructured film
at several T/T$_c$, where T$_c$ is the temperature ($\approx$ 11.8
K) at which zero-resistance state is reached. The magnetic field
applied in the plane of the film and orthogonal to current in
these measurements, was scanned in units of kOe following the
cycle 0 $\rightarrow$ 3.5 $\rightarrow$ 0 $\rightarrow$ - 3.5
$\rightarrow$ 0. A constant current density of 1$\times$10$^6$
A/cm$^2$, which corresponds to the J$_c$ at 9.8 K (T/T$_c$ = 0.83)
was used for resistance measurements. At small T/T$_c$ (= 0.79),
the R first remains zero with the increasing field. This is
expected as the zero-field current density J is smaller than the
J$_c$ at this T/T$_c$. However, as the field increases, J$_c$
drops and at H = H$^*$ (marked in the figure), the J becomes
larger than J$_c$ and the sample goes in a dissipative state. Here
the dependence of R is linear on H, as expected in the flux flow
regime of the mixed state. On reversing the field, however, the
resistance drops to zero much faster. The dissipation-less state
appears at a higher field H ($>$ H$^*$) in the reverse branch of
the loop. A similar hysteretic behavior is seen at the higher
values of T/T$_c$. In Fig. 3 (h) we show the result of a similar
measurement performed on an unpatterned bilayer. In this case the
R vs. H curve is completely reversible. It is clear from these
measurements that the hysteretic behavior of R vs. H is a
consequence of the nano-patterning. This is the central result of
our paper.

In Fig. 4(a) and (b) we summarize these observations of a
field-assisted reentrant superconducting state in terms of H-T
phase diagram. Fig. 4(c) shows a sketch of the R-H curve whose
critical points have been marked by letters A, B, C, D, E and F.
The point A corresponds to H$^*$ where superconductivity
disappears on ramping up the field, point B denotes the maximum
applied field and point C is where superconductivity reappears on
reducing the field from its maximum value. Similarly, points D and
F mark destruction and reappearance of SC on the reverse branch of
the loop while E corresponds to the maximum negative field. The
contours of the points C and D as a function of temperature are
shown in Fig. 4(a), and the shaded area enclosed by them is the
range of field where global superconductivity exists while
scanning it from + H$_{max}$ to - H$_{max}$. Fig. 4(b) is the
phase space of superconductivity for - H$_{max}$ to + H$_{max}$
excursion.

Before attempting to put forward a model which would allow us to
understand the results of Fig. 3 and 4, it is important to point
out that for our film of thickness d ($\approx$ 50 nm) much
smaller than the penetration depth $\lambda$ ($\approx$ 200 - 250
nm) placed in a parallel field, a large contribution to
dissipation may come from orbital and Pauli pair-breaking
processes, in addition to the dissipation due to flux flow.
However, the pair breaking processes are not hysteretic. The
misalignment of the magnetic field away from a true parallel
configuration is also an important issue because the parasitic
out-of-plane component of the field can lead to nucleation of
perpendicular vortices. To rule out this possibility, we had
aligned the field parallel to the plane of the film with a
precision of $\pm$ 0.1 degree using a specially built experimental
setup \cite{Patnaik}.


A correct model for understanding the phase diagram of Fig. 4 must
take into account two important features of the nano-structured
sample; first, it has a periodic granularity. The CoPt-NbN squares
become superconducting at $\approx$ 15.1 K, but a long-range phase
coherence develops only when the CoPt-free grids go
superconducting at $\approx$ 11.8 K. The patterned film therefore
is characterized by two distinct critical current densities, a
large J$_{cg}$ inside the NbN-CoPt squares and a much weaker
J$_{cj}$ in the channels separating these squares. Secondly, the
CoPt islands have in-plane $\vec{M}$ with coercive field $\approx$
250 Oe (see Fig. 2) which has been marked by a pair of horizontal
lines in the phase diagram of Fig. 4. It is clear that in most of
the phase space, the CoPt islands are saturated with their
$\vec{M}$ $\parallel$ $\vec{H}$. A qualitative understanding of
the flux distribution can be made from Fig. 5(a) and (c) drawn for
($\vec{H}$ $\perp$ $\vec{J}$) $\perp$ $\hat{n}$ and ($\vec{H}$
$\parallel$ $\vec{J}$) $\perp$ $\hat{n}$ configurations. We show
that even when CoPt squares are flux coupled at H $>$ H$_c$, some
fringing field from the edges of the squares produces flux lines
of opposite polarity in the NbN channels. These are marked by open
circles and crosses in the top view of the square array. The flux
due to external magnetic field has a different distribution in the
channels and the CoPt covered NbN squares because of their
different J$_c$'s. It can be understood in the framework of the
two-level critical state model of Ji, Rzchowski, Anand, and
Tinkham \cite{Ji}. Using this model we can define two types of
fluxons; (i) those pinned by pinning centers in the square
($\phi$$_p$), and (ii), the fluxons that are confined to the
channels and are free to move due to weak pinning ($\phi$$_f$). As
the external field is increased, fluxons enter the channels first,
supersaturating them and leading to the increase of dissipation
along A-B branch of the curve in Fig. 4(c). The flux eventually
enters the squares where it is pinned. On decreasing the field,
the free flux ($\phi$$_f$) leaks out from the channels to such an
extent that the zero resistance state is reestablished along the
low resistance paths marked in Fig. 5 at a higher value of the
external field. This is possible because there is a large pinned
flux ($\phi$$_p$) in the NbN squares.

We expect that in the $\vec{H}$ $\perp$ $\vec{J}$ configuration
(Fig. 5(a)), the current flow along the channels (paths which do
not cross CoPt-NbN squares) remain dissipative due to the dipolar
field of CoPt. This is where the magnetization of CoPt plays an
explicit role. A similar scenario also explains reentrant
superconductivity seen for the $\vec{H}$ $\parallel$ $\vec{J}$
configuration sketched at the left bottom panel (c) of Fig. 5. A
typical set of R vs. H data taken at T/T$_c$ = 0.85 in $\vec{H}$
$\parallel$ $\vec{J}$ configuration is shown in panel (d) of Fig.
5. The granular NbN channels although form a continuous path for
the flow of current for $\vec{H}$ $\parallel$ $\vec{J}$, the cores
of vortices, which are already large due to a small order
parameter in the channels, start overlapping, and even at low
fields, the channels become resistive. The onset of resistance
here has nothing to do with vortex motion under Lorentz force.
This type of force-free dissipation has been seen in granular
films of NbN \cite{Dhkim}.

The observations made in the present study have important
repercussions on the phenomenon of field assisted
superconductivity seen in FM-SC bilayers \cite{Yu}, trilayers
\cite{Gu} and dot structures \cite{Lange1}. While in the case of
FM-SC-FM trilayer structures placed in an in-plane field, no
inhomogeneity of the order parameter is introduced as long as each
FM layer behaves like a single domain film, the same is
unavoidable in bilayers at field H $<$ H$_c$ and in dot arrays
even beyond saturation. We believe that in such systems the
repercussions of the two-level critical state model warrant
consideration.

This research has been supported by a grant from the Nanoscience
$\&$ Nanotechnology Initiative of the Department of Science $\&$
Technology, Government of India, and internal funding of IIT
Kanpur. We thank Professor G. K. Mehta for his support in setting
up the FIB facility and to S. K. Bose, R. Sharma, and S.
Srivastava for technical assistance.

\newpage

\section*{FIGURE CAPTIONS:}

\noindent Fig. 1. A schematic of the sample design is shown in
panel (a). Panel (b) shows a scanning electron micrograph of the
patterned bilayer, and panels (c) and (d) AFM and MFM images
respectively. AFM and MFM measurements were performed
simultaneously on thermally demagnetized samples at lift height of
40 nm. AFM image of the sample is quite similar to its SEM image.
The bright and dark features in MFM represent domains of reversed
magnetization.

\noindent Fig. 2. R (T) of the patterned (filled circles) and
unpatterned (open circles) bilayers. A clear two-step
superconducting transition is seen in the patterned film. Inset
shows J$_c$ of the unpatterned (open circles) and patterned (solid
circles) samples. M vs. H loops for an unpatterned bilayer sample
at 5 K (open symbols) and 20 K (solid line) with field applied
parallel to the film surface are also shown in the inset. Here the
magnetization is measured in units of 10$^{-3}$ emu.

\noindent Fig. 3. R (H) of the nanoengineered CoPt-NbN/MgO bilayer
is shown in panel (a) to (g) at several T/T$_c$. Small arrows in
panel (e) mark the direction of field sweep. The critical field
H$^*$ at which dissipation appears on ramping the field from zero
has also been marked in the figures with vertical arrows. Panel
(h) shows R (H) of the unpatterned sample at T/T$_c$ = 0.89.

\noindent Fig. 4. The dark areas in panel (a) and (b) show the H-T
phase space where the FIS exists as shown by results of Fig. 3.
Panel (c) sketches the R-H curve in which critical points have
been marked by letters A, B, C, D, E and F. The contours of the
points C and D as a function of temperature are shown in (a) while
scanning the field from + H$_{max}$ to - H$_{max}$. Contours of
points F and A are shown in panel (b) for the field scan from -
H$_{max}$ to + H$_{max}$. The coercive field H$_c$ = $\pm$ 250 Oe
is drawn by a pair of horizontal lines in both the phase diagrams.

\noindent Fig. 5. Panel (a) Top view of flux distribution for
($\vec{H}$ $\perp$ $\vec{J}$)$\hat{n}$. In-plane field lines are
shown as dotted lines whereas the dipolar field penetrating the
NbN channel is shown by circles and crosses. Large arrows in the
middle of squares mark the direction of $\vec{M}$ of CoPt islands.
A side view of the dipolar field produced by the CoPt is sketched
in panel (b). Panel (c) shows the flux distribution when $\vec{H}$
and $\vec{J}$ are collinear. Panel (d) shows R vs. H plot in
$\vec{H}$ $\parallel$ $\vec{J}$ configuration at T/T$_c$ = 0.85.

\clearpage
\begin{figure}[h]
\vskip 0cm \abovecaptionskip 0cm
\includegraphics [width=12cm]{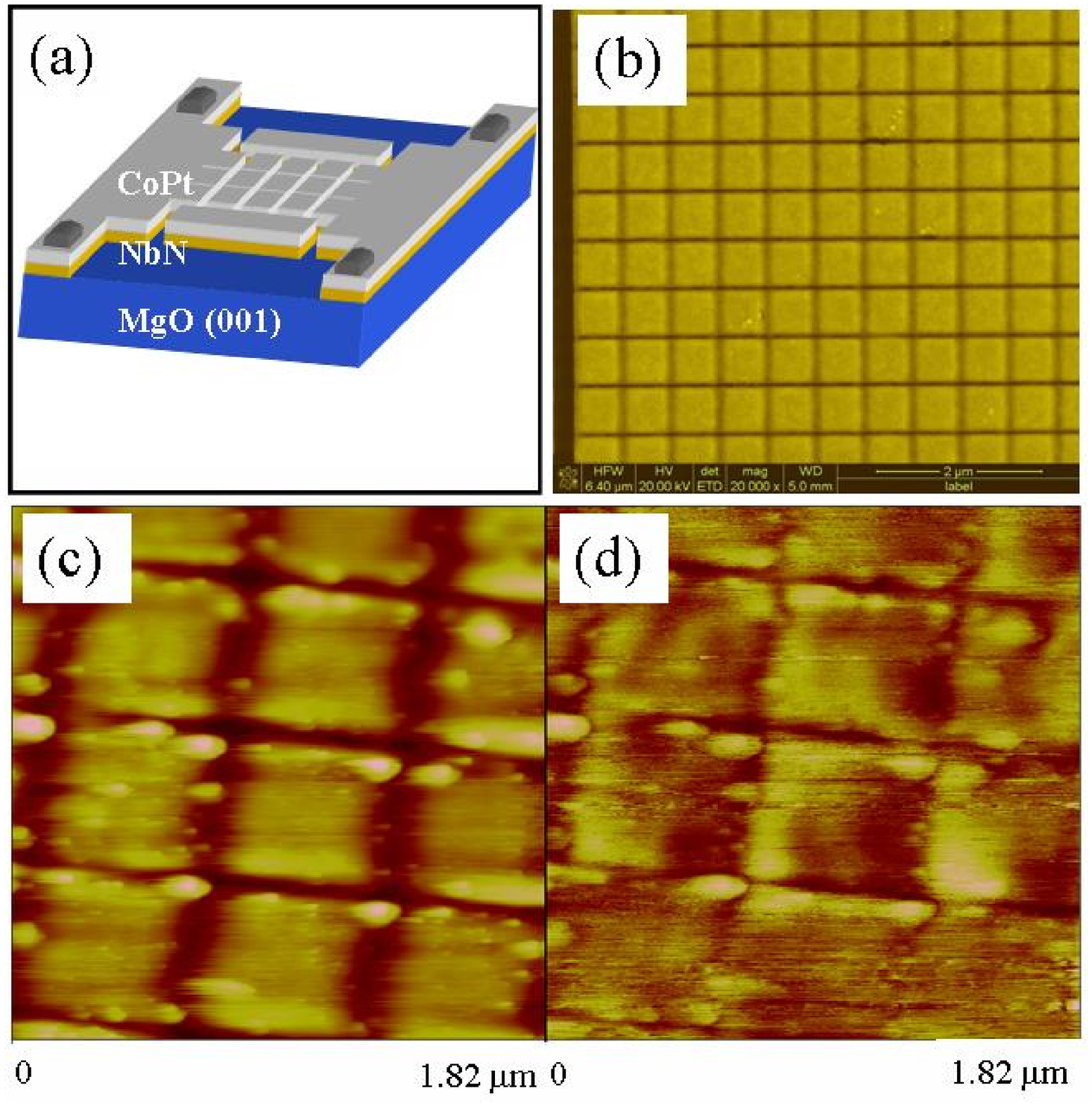}%
\caption{\label{fig1}}
\end{figure}

\clearpage
\begin{figure}[h]
\vskip 0cm \abovecaptionskip 0cm
\includegraphics [width=14cm]{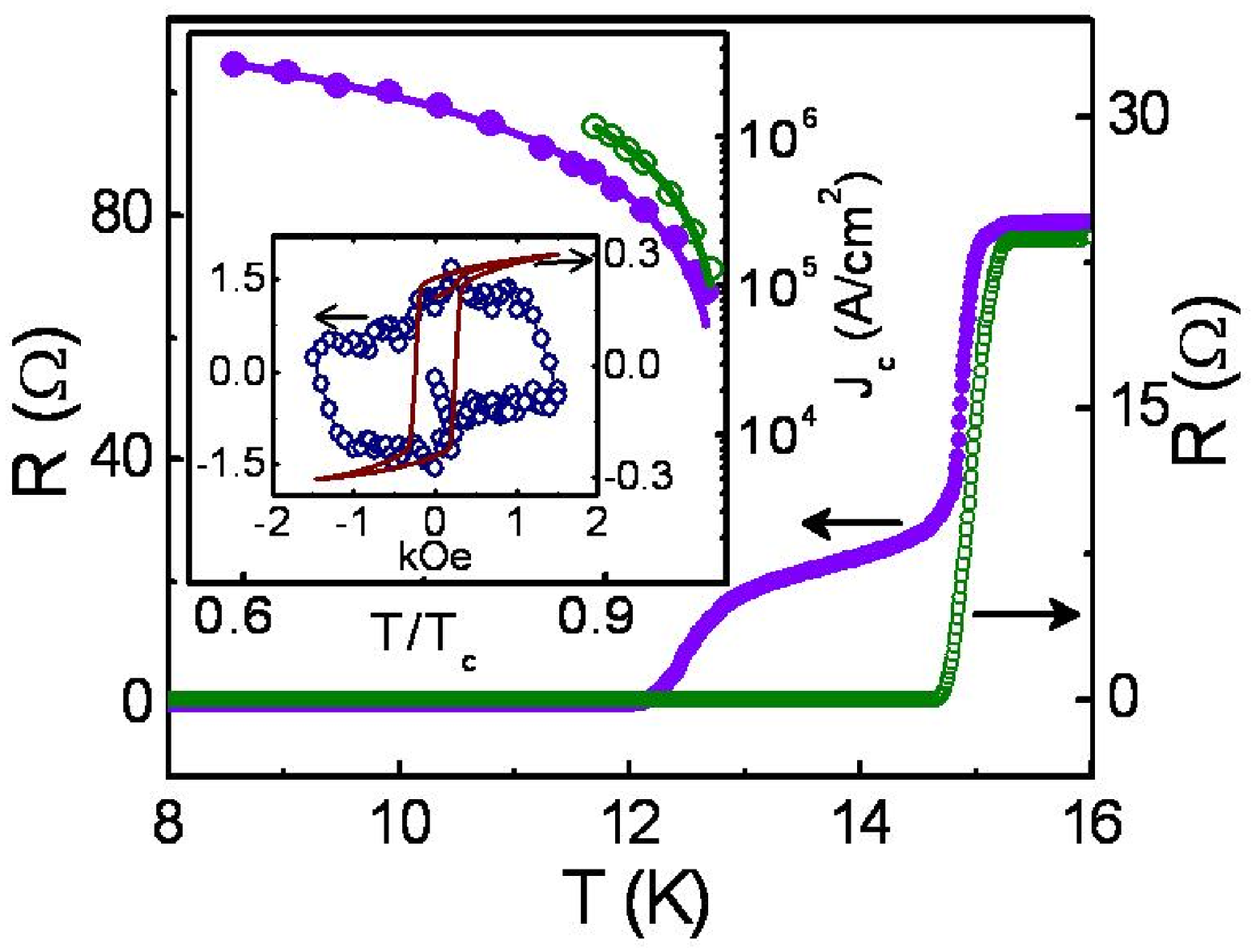}%
\caption{\label{fig2}}
\end{figure}

\clearpage
\begin{figure}[h]
\vskip 0cm \abovecaptionskip 0cm
\includegraphics [width=14cm]{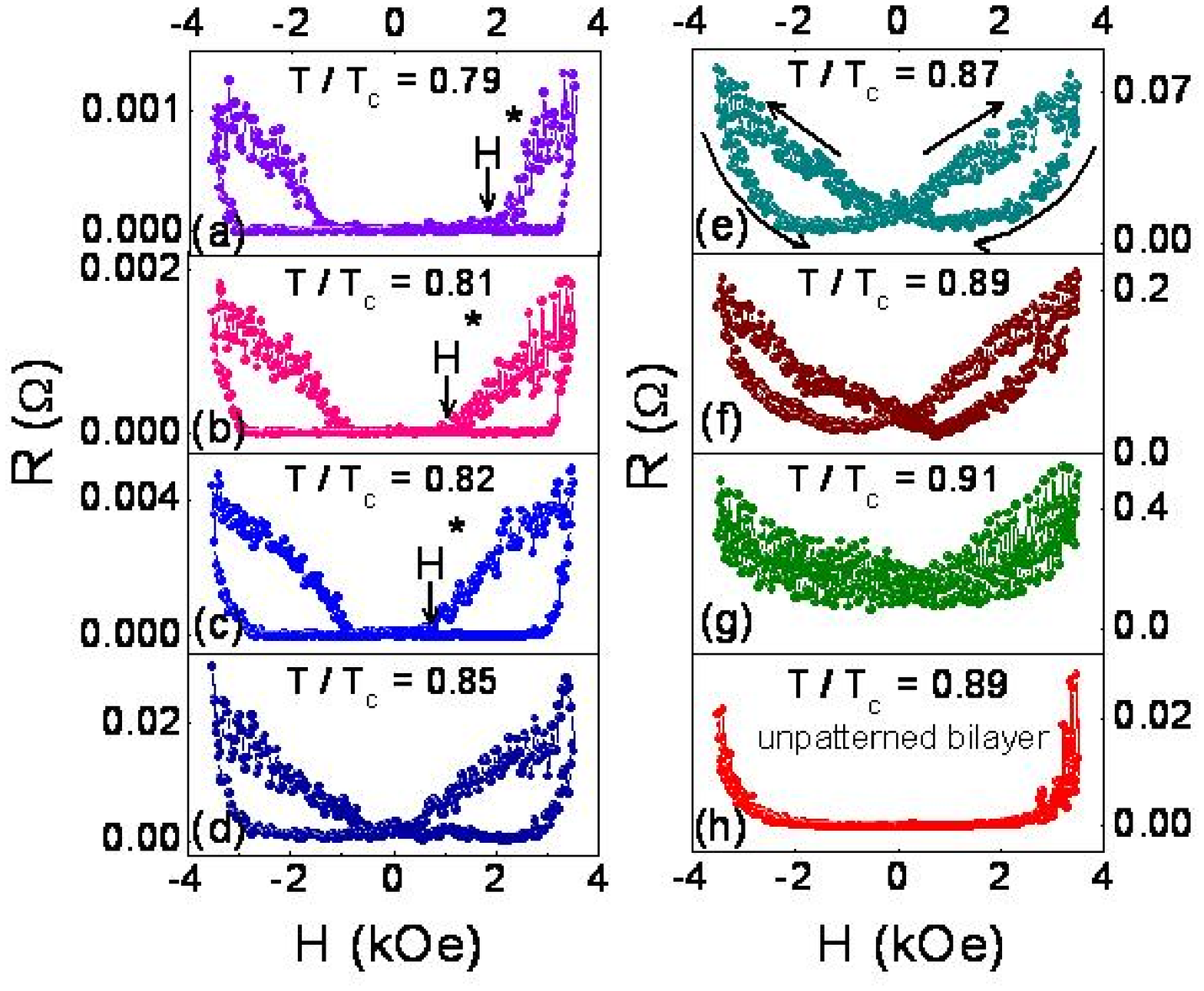}%
\caption{\label{fig3}}
\end{figure}

\clearpage
\begin{figure}[h]
\vskip 0cm \abovecaptionskip 0cm
\includegraphics [width=10cm]{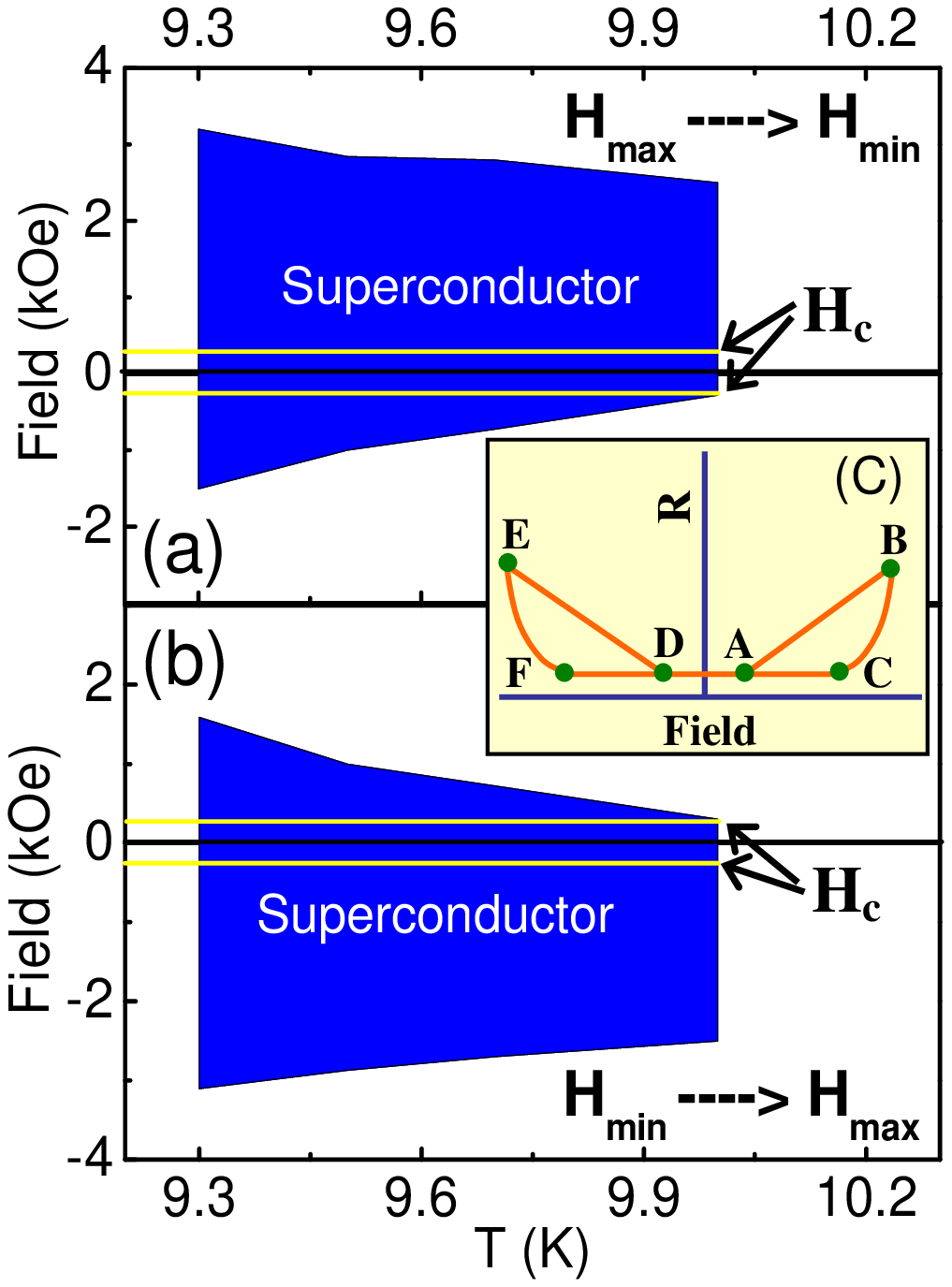}%
\caption{\label{fig4}}
\end{figure}

\clearpage
\begin{figure}[h]
\vskip 0cm \abovecaptionskip 0cm
\includegraphics [width=12cm]{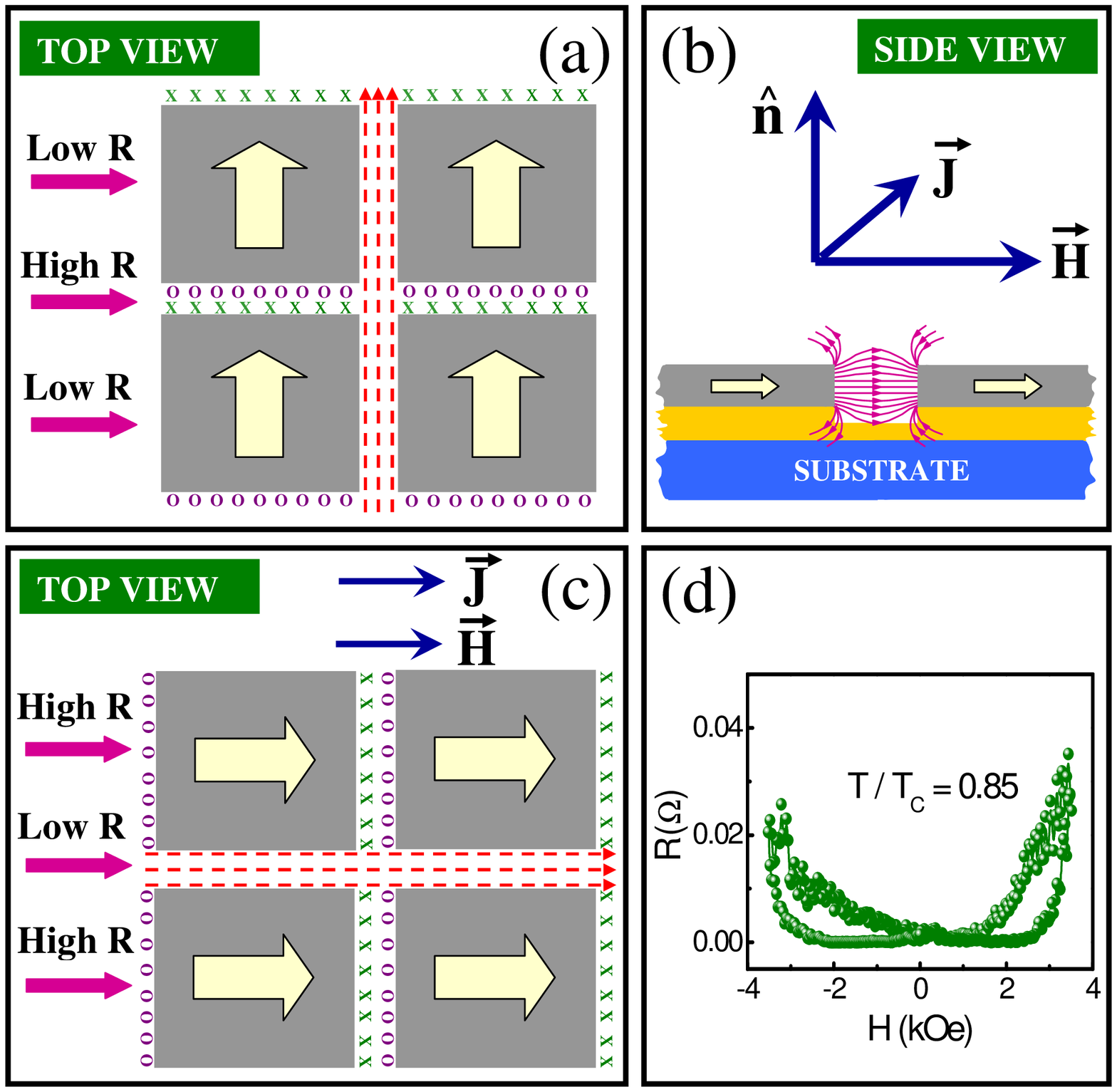}%
\caption{\label{fig5}}
\end{figure}

\end{document}